\documentclass{article}
\usepackage[numbers]{natbib}
\usepackage{amsmath}
\usepackage{tikz}
\usepackage{float}
\usetikzlibrary{shapes.geometric, arrows.meta, positioning, calc}
\usepackage{listings}
\usepackage{xcolor}

 \usepackage[preprint]{neurips_2025}

\usepackage[utf8]{inputenc} 
\usepackage[T1]{fontenc}    
\usepackage{hyperref}       
\usepackage{url}            
\usepackage{booktabs}       
\usepackage{amsfonts}       
\usepackage{nicefrac}       
\usepackage{microtype}      
\usepackage{xcolor}         

\title{CodeMem: Architecting Reproducible Agents via Dynamic MCP and Procedural Memory
}

\author{%
  Nishant Gaurav \\
  \texttt{nishant@agentr.dev} \\
  \And
  Adit Akarsh \\
  \texttt{adit@agentr.dev} \\
  \AND
  Tejas Ravishankar \\
  \texttt{tejas@agentr.dev} \\
  \And
  Manoj Bajaj \\
  \texttt{manoj@agentr.dev} \\
}

\begin{document}

\maketitle

\begin{abstract}
Current tool-using AI agents suffer from limited action space, context inefficiency, and probabilistic instability that makes them unsuitable for handling repetitive tasks which are otherwise reliably and efficiently tackled by agentic workflows built on platforms like n8n~\cite{n8n} and Zapier~\cite{zapier}. Earlier works like CodeAct~\cite{wang2024executable}, DynaSaur~\cite{nguyen2024dynasaur}, Code Mode~\cite{cloudflare_codemode} have tried to tackle the first two issues by using the whole Python language as its action space: The number of tools that the agent can call becomes infinite. Python code blocks can execute complex actions into a single step and print only relevant results which helps in keeping the context lean. However, the probabilistic instability issue still remains, as for the same task in the same environment, the agent can follow different trajectories due to the probabilistic nature of LLMs. Therefore, we need procedural memory for consistency and reliability. This paper proposes \textbf{CodeMem}, an architecture to implement procedural memory via code which can be used to build and run reusable agentic workflows with deterministic reliability.
\end{abstract}

\section{Introduction}
Tool-using language agents have evolved from single-shot chatbots into complex systems capable of planning and state management. Frameworks like CoALA emphasize that capable agents require structured memory, rich action spaces, and iterative decision-making loops~\cite{sumers2024coala}. However, most production architectures still rely on token-heavy, tool-centric interaction patterns where the LLM micromanages every step. This paper proposes CodeMem, an architecture that reframes the LLM as an architect of executable workflows. Instead of standard chat-based tool calling, the agent utilizes a sandbox to write, validate, and save successful logic into a persistent procedural memory bank. This approach solves the reproducibility crisis inherent in probabilistic models by shifting complex logic from volatile context windows into deterministic code.

The plan for the rest of the paper is as follows. We first summarize the relevant literature which form the basis for CodeMem (in Section~\ref{sec:related_work}). Then we dive deeper into the benefits of CodeAct over ReAct which establishes that Code is the right format for capturing procedures (in Section~\ref{sec:comparative_analysis}). The next Section covers the key bottlenecks which must be solved (in Section~\ref{sec:limitations}). Section~\ref{sec:codemem_architecture} proposes the CodeMem Architecture which overcomes the key bottlenecks specified in the previous section. In Section~\ref{sec:case_study}, we show how CodeMem creates procedural memory with a real-world case-study. Section~\ref{sec:experiments} shows experiments which prove the benefits of CodeMem quantitatively. Finally we summarize the findings in Section~\ref{sec:summary}.

\section{Related Work}
\label{sec:related_work}

\subsection{Procedural Memory in Theory}
The CoALA framework organizes agent memory into working, episodic, semantic, and procedural categories \citep{sumers2024coala}. While working memory (context) and semantic memory (RAG) are well-solved, procedural memory (implicit knowledge of how to execute tasks) remains a bottleneck. The paper suggests three possible approaches for capturing procedural memory in AI Agents:
\begin{enumerate}
    \item Rewriting the code of the Agent
    \item Editing the weights of the LLMs
    \item Editing the instructions of the AI Agent
\end{enumerate}

Theoretically both these approaches make sense, but practical implementations are still missing. LangGraph, a popular agentic framework, calls this out in their documentation:
\begin{quote}
``In practice, it is fairly uncommon for agents to modify their model weights or rewrite their code. However, it is more common for agents to modify their own prompts.'' \citep{langchain2024memory}
\end{quote}

\subsection{Implementing Procedural Memory via Meta Prompting}
Frameworks like LangGraph distinguish between short-term memory (thread-scoped checkpoints) and long-term memory (cross-thread persistence) \citep{langchain2024persistence}.

\paragraph{Short-Term:} LangGraph effectively manages conversation state (i.e. the messages history, including human, AI, and tool messages) via checkpoints. All kinds of memory (semantic, episodic, and procedural) can be captured in short term memory but there is no guarantee that the agent will utilize this memory reliably. That’s why we need long term memory.

\paragraph{Long-Term:} For capturing semantic and episodic memory we can leverage RAG and there are several open-source projects (e.g. Graphiti \citep{graphiti_github}, LightRAG \citep{lightrag_github}) which are getting used in production. However, capturing procedural memory is still an underexplored topic. LangGraph conceptualizes procedural updates primarily as modifications to graph topology or just the system instructions (e.g., ``update the prompt to handle X better next time'') \citep{langchain2024memory}.

For example, the LangGraph team built a Tweet generator using external feedback and prompt re-writing to produce high-quality paper summaries for Twitter. In this case, the specific summarization prompt was difficult to specify \textit{a priori}, but it was fairly easy for a user to critique the generated Tweets and provide feedback on how to improve the summarization process \citep{langchain_memory_concepts}.

\begin{figure}[hbt!]
    \centering
    \includegraphics[width=0.5\textwidth]{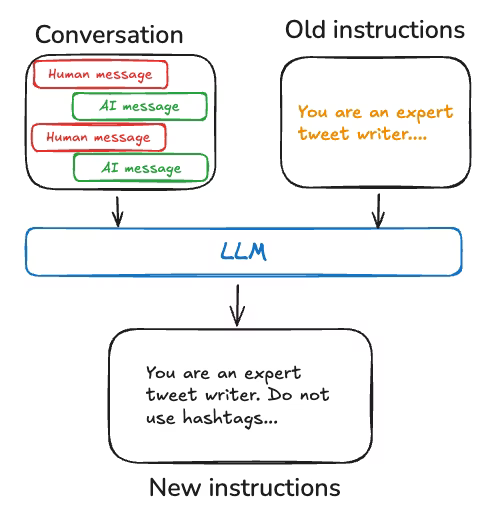}
    \caption{Mechanism for updating procedural memory via instruction tuning. The agent analyzes the conversation history and existing instructions to generate refined new instructions (e.g., adding constraints like "Do not use hashtags"). This process, utilized by the LangGraph Tweet generator, relies on meta-prompting rather than code modification \citep{langchain_memory_concepts}.}
    \label{fig:langgraph_memory}
\end{figure}

However, relying on prompt updates for procedural memory is brittle. Because LLMs are stochastic, updating a system prompt does not guarantee the agent will adhere to the new procedure in the future, especially for longer instructions.

In contrast, \textbf{CodeMem} treats procedural memory as frozen code, ensuring deterministic execution without relying on the model’s instruction-following capability.

\subsection{Progressive Disclosure for Tool Calling}
Standard approaches to tool use require injecting full tool definitions into the system prompt. As tool libraries grow, this leads to context window bloat and attention degradation. Dynamic ReAct \citep{gaurav2025dynamic} proposes a solution by decoupling tool existence from tool definition. Instead of loading all tools, the agent is given a discovery mechanism to query a registry and load only relevant tools and their schemas on-demand.

This follows the principle of progressive disclosure where we disclose the relevant tools to the LLM on an ad hoc basis. This shifts tool access from an $O(N)$ context cost to an $O(1)$ search operation, enabling infinite-scale tool libraries. This approach was also later highlighted in Anthropic’s engineering blog titled \textit{Code execution with MCP} \citep{anthropic_mcp_code}.

Access to a tool registry is also a key ingredient for an infinite action space. Without it, a CodeAct Agent cannot leverage capabilities outside of first party python libraries.

\section{Comparative Analysis: Executable Code vs. Direct Tool Calling}
\label{sec:comparative_analysis}

Recent benchmarks indicate a fundamental limit to the JSON-centric tool calling paradigm (ReAct), where LLMs are provided a set of tools (with tool names and tool arguments), and they write JSON text for calling these tools. By shifting from discrete tool calls to executable code, agents achieve higher success rates and lower latency.

\subsection{The Performance Gap: CodeAct vs. ReAct}
Research by \citet{wang2024executable} on the CodeAct framework demonstrates that agents utilizing executable Python code as their primary action space significantly outperform those relying on standard text-based or JSON-based tool invocation.

\begin{itemize}
    \item \textbf{Success Rate:} On the M3ToolEval benchmark, code-driven agents achieved a 20\% higher success rate compared to ReAct agents.
    \item \textbf{Action Efficiency:} Code agents required 30\% fewer turns to complete equivalent tasks. This efficiency stems from the ability to compose multiple actions (e.g., search $\rightarrow$ filter $\rightarrow$ sort) into a single executable block, whereas ReAct requires a full LLM inference cycle for each individual step.
\end{itemize}

\subsection{Decoupling Reasoning from Computation}
Standard agents suffer from ``computation hallucination,'' where the LLM attempts to perform arithmetic or logic within the context window. \citet{chen2022program} introduced Program of Thoughts (PoT), showing that offloading logic to a Python interpreter improves numerical reasoning accuracy by over 10\% compared to Chain of Thought (CoT). In the CodeMem architecture, the Dynamic MCP tools serve as the ``API,'' but the logic (loops, data processing) is offloaded to the Python Sandbox. This separation ensures that the LLM acts as a semantic router while the CPU handles the syntactic execution, eliminating non-deterministic errors in logical processing.

\subsection{Contextual Complexity Analysis}
We define the context cost $C$ for a workflow of $N$ steps.

\begin{itemize}
    \item \textbf{Standard Tool Calling:} 
    \[ C \approx \sum_{i=1}^{N} (S_{\text{prompt}} + S_{\text{history}} + S_{\text{tool\_output}_i}) \]
    The context grows linearly with every intermediate step, often truncating memory before the task is complete.
    
    \item \textbf{CodeMem:} 
    \[ C \approx S_{\text{prompt}} + S_{\text{code\_block}} + S_{\text{final\_result}} \]
    Intermediate data transformations occur in the sandbox’s memory (RAM), not the LLM’s context window. This allows for infinite-depth processing (e.g., looping through 10,000 database rows) without consuming token budget.
\end{itemize}

\begin{table}[hbt!]
\centering
\caption{Benchmark Comparison: JSON Tool Calling vs. CodeMem Execution}
\label{tab:benchmark_comparison}
\begin{tabular}{|l|l|l|}
\hline
\textbf{Metric} & \textbf{Standard Tool Calling (JSON)} & \textbf{CodeMem (Python)} \\ \hline
Multi-step Latency & High ($N \times$ Inference Time) & Low ($1 \times$ Inference + Execution Time) \\ \hline
Logic Reliability & Probabilistic (LLM predicts logic) & Deterministic (Python executes logic) \\ \hline
Max Loop Depth & $< 25$ (Limited by Context) & $\infty$ (Limited by Compute/Time) \\ \hline
Reproducibility & Low (Varies by seed/drift) & High (Versioned Scripts) \\ \hline
\end{tabular}
\end{table}

\section{The Limitations of Vanilla CodeAct}
\label{sec:limitations}

While shifting from ReAct to CodeAct addresses the issues of context inefficiency and limited action spaces, it introduces a new set of bottlenecks related to reliability and persistence. CodeAct allows an agent to utilize the full expressivity of Python \citep{wang2024executable}, but without a dedicated memory architecture, it suffers from issues arising from the nature of LLMs. We identify four structural penalties of using vanilla CodeAct without a persistence layer:

\subsection{Probabilistic Instability}
Standard CodeAct relies on generative improvisation for every task. Because LLMs are non-deterministic, the agent may generate different code paths for the exact same request across different sessions, especially if the request is not detailed (e.g., using \texttt{pandas} one day and \texttt{openpyxl} the next). This leads to \textit{trajectory divergence}, where the logic validated in a previous session is not guaranteed to be reproduced, effectively causing the agent to ``reinvent the wheel'' unpredictably.

\subsection{Redundant Computation and Latency}
Re-deriving solutions from scratch consumes extra tokens and requires computational time, even if provided perfect instructions. The total time for a task is calculated as:
\[
T_{\text{task}} = T_{\text{plan}} + T_{\text{write\_code}} + T_{\text{debug}} + T_{\text{execute}}
\]
Even for tasks solved hundreds of times, the agent wastes tokens and time re-generating the planning and coding phases, since in practice it usually does not get everything right in one go. A system with procedural memory avoids this overhead by reducing the cost to approximately $T_{\text{execute}}$.

\subsection{Inability to Retain Feedback}
Vanilla CodeAct agents lack cross-session memory, making them unsuitable for workflows with any customization. If a user provides specific feedback, such as ``filter by `received' date rather than `sent' date'', that correction is lost once the current context window closes. Without a mechanism to save successful logic, the agent fails to learn from the environment and requires repeated supervision for the same errors.

\subsection{Context Drift}
In longer, multi-step tasks (e.g., ``Scrape 100 pages, filter by keyword, then summarize''), standard CodeAct agents often suffer from context drift. As the conversation history fills with intermediate code outputs and error logs, the original high-level goal gets pushed out of the LLM’s attention span. The agent might successfully scrape the data but ``forget'' to summarize it, or it might get stuck in a loop trying to fix a minor scraping error, losing sight of the overall objective.

\section{The Solution: CodeMem Architecture}
\label{sec:codemem_architecture}

To overcome the structural penalties of vanilla CodeAct, we propose CodeMem, an architecture that transitions the LLM from a perpetual improviser to an architect of reusable skills. We address the bottlenecks identified in Section 4 by combining a Dynamic MCP discovery layer with a Procedural Memory Bank.

\subsection{The Core Toolset}
The foundation of the architecture relies on four primary tools that enable the agent to navigate, plan, and execute logic deterministically:

\begin{itemize}
    \item \texttt{search\_functions} \& \texttt{load\_functions}: Implements the Dynamic MCP protocol \citep{gaurav2025dynamic}. This allows the agent to discover relevant tools and inject their schemas Just-in-Time, preventing context window bloat.
    
    \item \texttt{write\_todos}: A state persistence mechanism for long-horizon planning. It explicitly tracks dependencies (Pending $\rightarrow$ In Progress $\rightarrow$ Completed), preventing the ``Goal Drift'' often seen in standard CodeAct sessions.
    
    \item \texttt{execute\_code}: Runs Python scripts in a secure sandbox \citep{anthropic2025skills}, allowing the agent to chain loaded tools together programmatically rather than sequentially via chat.
    
    \item \texttt{register\_skill}: The tool to add to procedural memory. This tool takes a successfully validated Python function and saves it into the user's permanent library. It freezes the logic (and any specific user constraints) after it is finalized so it can be recalled deterministically in future sessions without re-generation.
\end{itemize}

\subsection{Solving Instability}
To resolve Probabilistic Instability (Section 4.1) and the lack of a method to incorporate user feedback (Section 4.3), CodeMem changes the lifecycle of code generation. Instead of treating code as ephemeral text, we treat it as a candidate for permanent storage.

\begin{description}
    \item[Trial:] The agent uses \texttt{execute\_code} to generate and test logic in the sandbox, self-correcting errors in real-time.
    \item[Saving:] Once the output matches user requirements including any specific constraints like ``filter by received date'', the system triggers \texttt{register\_skill}.
    \item[Persistence:] The validated logic is saved to the user's library, and used deterministically in the future.
\end{description}

\subsection{Solving Latency and Cost}
To eliminate the Re-Derivation Penalty (Section 4.2), CodeMem alters the execution flow for recurrent tasks.

\begin{itemize}
    \item \textbf{Standard CodeAct:} 
    \[ T_{\text{task}} = T_{\text{plan}} + T_{\text{write}} + T_{\text{debug}} + T_{\text{execute}} \]
    
    \item \textbf{CodeMem Reuse:} 
    \[ T_{\text{task}} \approx T_{\text{execute}} \]
\end{itemize}

The user is able to access their repository of stored skills, and load it into the execution environment’s context as well as the LLM’s context. This allows the LLM to be aware of the function’s functionality, while also not having to rewrite the code to use it. Thus, by skipping the planning, writing/rewriting, and debugging phases entirely, the agent reduces latency and eliminates the token costs associated with re-generation.

\subsection{Solving Context Drift}
To prevent context drift (Section 4.4) during longer tasks, CodeMem utilizes \texttt{write\_todos} as an external anchor.
Implicit internal reasoning often fails as the context window fills with logs and code blocks. \texttt{write\_todos} maintains a structured checklist (Pending $\rightarrow$ In Progress $\rightarrow$ Completed) outside of the chat history. This allows the agent to deep-dive into specific sub-tasks (like debugging a scraper) without losing its place in the overall plan, ensuring the final crystallized code is complete and modular.

\begin{figure}[ht]
    \centering
    \begin{tikzpicture}[
    scale=0.8, 
    transform shape,
    node distance=1.8cm and 2.0cm,
    auto,
    process/.style={rectangle, draw=black, fill=blue!5, thick, text width=2.2cm, align=center, rounded corners, minimum height=1.0cm},
    decision/.style={diamond, draw=black, fill=orange!10, thick, text width=1.2cm, align=center, aspect=2},
    database/.style={cylinder, shape border rotate=90, draw=black, fill=green!5, thick, aspect=0.25, text width=1.6cm, align=center, minimum height=1.5cm},
    endpoint/.style={ellipse, draw=black, fill=gray!10, thick, text width=2.0cm, align=center, minimum height=0.8cm},
    arrow/.style={-Latex, thick, draw=black!80},
    line/.style={thick, draw=black!80}
]

        
        \node (user) [endpoint] {User Query};
        \node (agent) [process, below=of user, yshift=0.5cm] {\textbf{Agent (LLM)}\\Architect};
        \node (sandbox) [process, below=of agent] {\textbf{Python Sandbox}\\\texttt{execute\_code}};
        \node (check) [decision, below=of sandbox] {Success?};
        \node (output) [endpoint, below=of check, yshift=-0.5cm] {Final Output};

        \node (registry) [database, left=of agent, xshift=-1cm] {\textbf{Tool Registry}\\(Dynamic MCP)};
        \node (memory) [database, right=of agent, xshift=1cm] {\textbf{Procedural Memory}\\(Saved Skills)};


        \draw [arrow] (user) -- (agent);
        \draw [arrow] (agent) -- node [right, font=\small, text width=2cm] {1. Gen Script} (sandbox);
        \draw [arrow] (sandbox) -- (check);
        \draw [arrow] (check.south) -- node [right, font=\small] {Result} (output);

        \draw [arrow, bend right=25] (agent.160) to node [midway, above, font=\small] {search} (registry.north);
        \draw [arrow, bend right=25] (registry.south) to node [midway, below, font=\small] {schemas} (agent.200);

        \draw [arrow, bend left=25] (agent.20) to node [midway, above, font=\small] {recall} (memory.north);
        \draw [arrow, bend left=25] (memory.190) to node [midway, below, font=\small] {functions} (agent.340);

        \draw [line] (check.west) -- node [midway, above, font=\small] {No} ++(-1.5, 0) coordinate (turn_err);
        \draw [arrow] (turn_err) |- node [near end, above, font=\small, rotate=90] {Retry (Stderr)} (agent.west);

        \draw [line] (check.east) -- node [midway, above, font=\small] {Yes} ++(1.5, 0) coordinate (turn_succ);
        
\draw [arrow] (turn_succ) -| node [near end, xshift=-0.2cm, font=\small, rotate=90] {\texttt{register\_skill}} (memory.south);

    \end{tikzpicture}
    \caption{\textbf{The CodeMem Architecture.} The Agent creates reproducible workflows by discovering tools (Left), executing them in a sandbox (Center), and freezing successful logic into persistent memory (Right).}
    \label{fig:codemode_arch}
\end{figure}
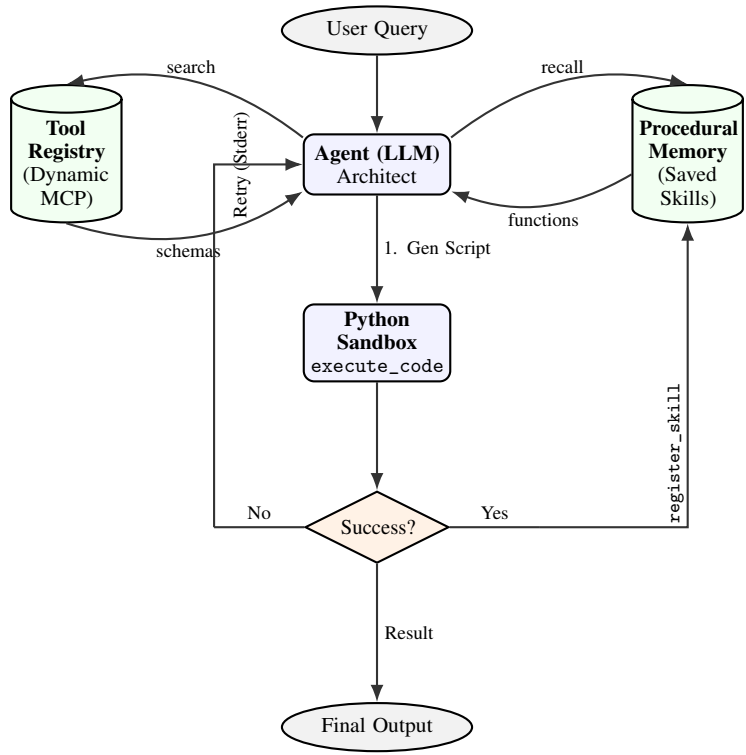

\section{Creating Procedural Memory: A Case Study}
\label{sec:case_study}

The ultimate goal of the CodeMem architecture is not merely to execute a task once, but to crystallize the successful logic into a reusable skill. This process transforms ephemeral chains of thought into persistent procedural memory. To illustrate the necessity and mechanics of this transition, we examine the \textbf{Outlook-OneDrive Bridge}, a complex administrative task that requires temporal filtering, conditional logic, and cross-service file orchestration.

\subsection{The Challenge: High-Fidelity Data Orchestration}
Consider the following real-world request, which represents a class of tasks typically fatal to standard chat agents:

\begin{quote}
\textbf{User Query:} ``Go through the past 15 days of emails in my Outlook. Wherever there’s a PDF or XLSX attachment, save it to a OneDrive folder named `Email Attachments/[Company Name]'. Ignore internal emails from @agentr.dev. If the company name is codeword, extract the real company name from the attachment metadata.''
\end{quote}

This prompt contains four distinct constraints: a temporal window (15 days), a file-type filter (PDF/XLSX), a negative filter (ignore domain), and a conditional logic branch (pseudo-code extraction).

\subsubsection{The ReAct Failure Mode: Context Collapse}
In a standard Tool-Calling (ReAct) architecture, the agent attempts to solve this iteratively. This approach fails due to linear complexity scaling:

\begin{enumerate}
    \item \textbf{Ingestion:} The agent calls \texttt{outlook\_list\_emails} and receives a JSON list of 50+ emails.
    \item \textbf{Iteration:} It must iterate through this list one by one. For every email with an attachment, it calls \texttt{outlook\_get\_attachment}.
    \item \textbf{Collapse:} The API returns the raw Base64 string of a 5MB PDF. This single data payload instantly fills the context window. The agent undergoes \textit{Context Collapse}—the truncation of the conversation history deletes the memory of which emails have already been processed. The agent typically enters a failure loop, re-fetching the first email indefinitely until the token budget is exhausted.
\end{enumerate}

\subsection{The CodeMem Solution: Sandbox Orchestration}
In the CodeMem architecture, the LLM does not ingest the data; instead, it architects the pipeline that moves the data. The agent operates in three distinct phases to ensure reliability.

\begin{description}
    \item[Phase 1: Discovery \& Planning.] The agent utilizes \texttt{search\_functions} to map the API surface, identifying the necessary tools for retrieval and storage.
    \item[Phase 2: Script Generation.] Instead of calling tools one by one, the agent authors a robust Python script (Listing \ref{lst:orchestration_script}) that handles the logic \textit{in silico}.
    \item[Phase 3: Execution.] The script executes in the sandbox. In our experiments, it processed 7 emails, filtered 3, and uploaded 4 files in a single execution pass taking 14 seconds. The final artifact is the \texttt{agent\_main} function itself, which is saved to the user’s library for zero-shot reuse.
\end{description}

\subsection{Architectural Advantages}
The shift from ReAct to CodeMem introduces two fundamental improvements in agent reliability:

\begin{itemize}
    \item \textbf{$O(1)$ Context Complexity:} In ReAct, context consumption scales linearly with the size of the data ($O(D)$), as file contents must pass through the LLM. In CodeMem, context consumption is constant ($O(1)$) relative to data size. The LLM only sees the reference to the data (e.g., variable \texttt{attachment['content']}), while the actual binary transfer occurs within the Python Sandbox’s heap memory. This allows agents to process gigabytes of data without utilizing a single token of context.
    
    \item \textbf{Deterministic Control Flow:} The requirement to ``ignore emails from @agentr.dev'' is a simple boolean check. In an LLM-driven loop, the model must ``remember'' to apply this rule at every step, making it susceptible to attention drift. In the generated script, the \texttt{if} statement is syntactically enforced by the Python interpreter, ensuring 100\% compliance with the negative filter.
\end{itemize}
\newpage
\begin{lstlisting}[language=Python, caption=The Agent-Generated Orchestration Script, label=lst:orchestration_script]
async def agent_main(days_back=15):
    # 1. Deterministic Date Calculation
    cutoff = datetime.now() - timedelta(days=days_back)
    filter_query = f"receivedDateTime >= {cutoff} and hasAttachments eq true"

    # 2. Bulk Fetch & Local Filter
    emails = await outlook__list_emails(filter=filter_query)

    for email in emails:
        # Logic handled in loop, not via tokens
        if "@agentr.dev" in email['from']: 
            continue

        # 3. Data Streaming (Not Context Loading)
        attachment = await outlook__get_attachment(email['id'])

        # 4. Metadata Extraction via Helper
        meta = await _analyze_email_metadata(email)
        path = f"Email Attachments December/{meta['company']}/"
        
        await onedrive__upload_file(path, attachment['content'])
\end{lstlisting}

\subsection{Pathways to Formation: Instruction vs. Exploration}
There are two distinct ways to turn a one-time conversation into a reusable tool. The reliability of the resulting tool depends on whether the user provides the blueprint (Instruction) or the agent discovers it (Exploration).

\subsubsection{Pathway 1: Exploration-Driven (The Scientist)}
In this scenario, the user provides a high-level goal (``Fetch my emails...'') without implementation details. The agent must perform Hierarchical Task Analysis, decomposing the abstract request into executable steps through trial and error.

\begin{itemize}
    \item \textbf{Cognitive Load: High.} The agent must hypothesize which tools to use (e.g., searching for ``fetch emails''), discover that the API lacks specific filters (e.g., filtering by attachment type), and devise a Python-side logic to compensate.
    \item \textbf{Execution Profile: Iterative and Cyclic.} The logs show the agent loading tools, attempting execution, discovering limitations (e.g., lack of file writing capabilities), and pivoting strategies (e.g., switching from \texttt{markitdown} to \texttt{pandas}).
    \item \textbf{Risk:} Without a clear map, the agent is prone to \textit{Goal Drift}—forgetting the high-level objective while fixing low-level syntax errors.
\end{itemize}

\subsubsection{Pathway 2: Instruction-Driven (The Architect)}
In this scenario, the user acts as the Architect, providing a fully specified algorithm. The agent acts as a Builder, translating natural language constraints directly into control flow. This pathway is characterized by a ``Handshake'' protocol where the agent confirms the architectural details before building.

\begin{quote}
\textbf{User Prompt:} ``Create a reproducible workflow... [Detailed logic for Outlook/OneDrive]... Ignore emails where the domain is @agentr.dev... Set an appropriate name...''
\end{quote}

Because the instructions are dense, the agent recognizes the need for state management and halts to verify requirements before writing the final code:

\begin{description}
    \item[Agent:] ``Since this workflow fetches emails from a rolling window... running it multiple times could result in duplicate files. Would you like me to add a `memory' to this agent (e.g., using a Google Sheet) to track processed emails?''
    \item[User:] ``no log required''
    \item[Agent:] [Generates Plan] ``I will filter out emails from @agentr.dev... and use the current month for dynamic folder naming.''
    \item[User:] ``ok''
\end{description}

This confirmation step is critical. It ensures that the resulting procedural memory aligns exactly with user intent before the logic is crystallized into the codebase.

\subsection{State Anchoring via \texttt{write\_todos}}
To manage the complexity of both pathways, particularly the exploration-driven one, the agent utilizes \texttt{write\_todos} as an External Working Memory. Before executing complex logic, the agent externalizes its plan:

\begin{verbatim}
todos:
  - status: completed
    content: Load functions for Outlook and OneDrive
  - status: pending
    content: Fetch Outlook emails... filter for (.pdf, .xlsx)
  - status: pending
    content: Process sample... extract Company Name...
  - status: pending
    content: Create folders... and upload
\end{verbatim}

This bridges the gap between Goal and Execution. If the sandbox crashes or the context is truncated due to a large error log, the agent reads the \texttt{todos} object to perform \textit{State Recovery}. It identifies that ``Process sample'' is still marked pending and immediately generates a fix for the error, picking up exactly where it left off without needing to re-derive the plan from the chat history.
\section{Experimental Evaluation}
\label{sec:experiments}

\subsection{Dataset Composition}
To rigorously evaluate the efficacy of the CodeMem architecture, we curated a dataset of 25 multi-step agentic tasks. These tasks range from simple retrieval (Difficulty 1) to complex, multi-tool orchestration (Difficulty 5). The dataset covers diverse domains including:
\begin{itemize}
    \item \textbf{Personal Productivity:} "Find unread emails from the last 24 hours and draft replies."
    \item \textbf{Data Analysis:} "Analyze Census data for ZIP 94012 and email a visualized report."
    \item \textbf{System Operations:} "Scrape Reddit threads, filter by sentiment, and update a Google Sheet."
\end{itemize}
Each task requires the agent to navigate a real-world environment, chaining between 2 to 8 distinct tools (e.g., Gmail, Google Sheets, Reddit, Web Search) to achieve a verifiable outcome.

\subsection{Evaluation Methodology}
Given the open-ended nature of agentic tasks, we employed an \textbf{LLM-as-a-Judge} framework to measure correctness. We utilized Gemini 2.5 Flash as the evaluator model.

Crucially, the judge was not limited to checking the final text response. It was granted access to the full agent trajectory, including:
\begin{enumerate}
    \item The sequence of tool selections and arguments.
    \item The raw code executed in the sandbox.
    \item The execution outputs (stdout/stderr) and side effects (e.g., API success responses).
\end{enumerate}
This allowed the judge to verify functional success (e.g., "Did the file actually get saved to Drive?") rather than relying solely on the agent's self-reported success.

\subsection{CodeMem Performance Analysis}
We evaluated the CodeMem architecture across three state-of-the-art foundational models. Table \ref{tab:codemem_perf} presents the comparative performance metrics.

\begin{table}[hbt!]
    \centering
    \caption{Performance Comparison on CodeMem Benchmarks}
    \label{tab:codemem_perf}
    \begin{tabular}{@{} l c c c c @{}}
        \toprule
        \textbf{Model} & \textbf{Correctness (Min)} & \textbf{Avg. Calls} & \textbf{P50 Latency} & \textbf{Total Tokens} \\
        \midrule
        Gemini 3 Full & \textbf{96\%} & 7.00 & 100.48s & 2.02M \\
        Claude 4.5 Sonnet & 79\% & 6.96 & 71.38s & 2.61M \\
        GPT-5 Chat & 68\% & 2.80 & \textit{14.75s} & \textit{0.49M} \\
        \bottomrule
    \end{tabular}
\end{table}

As shown in Table \ref{tab:codemem_perf}, \textbf{Gemini 3 Full} achieved the highest stability with a minimum correctness of 96\%, utilizing an average of 7.00 assistant calls to fully resolve tasks. Conversely, while \textbf{GPT-5 Chat} appears highly efficient in terms of latency (14.75s) and token usage (0.49M), this efficiency is illusory. The model frequently failed to iterate on edge cases, exiting the loop prematurely (2.80 avg calls) compared to the $\sim$7 calls required by higher-reasoning models to achieve success.

\section{Summary of Architecture}
\label{sec:summary}
\begin{table}[H]
    \centering
    \caption{Comparison of Architectural Components and Problems Solved}
    \label{tab:architecture_summary}
    \begin{tabular}{@{} l l p{6.5cm} @{}}
        \toprule
        \textbf{Feature} & \textbf{Implementation Tool} & \textbf{Problem Solved} \\
        \midrule
        Context Management & \texttt{search} / \texttt{load\_functions} & \textbf{Context Bloat:} Tools are loaded Just-in-Time via Dynamic MCP~\cite{gaurav2025dynamic}. \\
        \addlinespace
        State Persistence & \texttt{write\_todos} & \textbf{Goal Drift:} Complex plans are tracked in a structured list, preventing agents from "forgetting" parallel tasks. \\
        \addlinespace
        Flow Control & \texttt{execute\_code} (Sandbox) & \textbf{Rigid Orchestration:} Loops and conditionals occur in deterministic code, not probabilistic chat. \\
        \addlinespace
        Computation & \texttt{execute\_code} (Pandas) & \textbf{Logic Failure:} Mathematical operations are calculated by the CPU, not predicted by tokens. \\
        \addlinespace
        Reliability & \texttt{register\_skill} (DB) & \textbf{Reproducibility:} Successful logic is frozen into reusable skills, avoiding stochastic prompt drift. \\
        \bottomrule
    \end{tabular}
\end{table}

\bibliography{main}

\end{document}